\documentclass[a4paper]{jpconf}
\usepackage{amsmath,amssymb,bbm}
\usepackage{amsfonts}
\usepackage{graphicx}
\usepackage{pgf}

\newlength{\figurewidth}
\newcommand\Schr{Schr\"odinger}
\usepackage{amsmath,amssymb,bbm}
\usepackage{amsfonts}
\usepackage{graphicx}
\usepackage{pgf}
\usepackage{multirow}

\newcommand\ro{\widehat\rho}
\newcommand\Ho{\widehat H}
\newcommand\Qo{\widehat Q}
\newcommand\fo{\widehat f}

\begin{document}
\title{The case of Quantum Gravity with Spontaneous Collapse of the Wave Function}

\author{Lajos Di\'osi}

\address{Wigner Research Centre for Physics, H-1525 Budapest 114, P.O.Box 49, Hungary\\
E\"otv\"os Lor\'and University, H-1117 Budapest, P\'azm\'any P\'eter stny. 1/A}

\ead{diosi.lajos@wigner.hu}

\begin{abstract}
When about half a century ago the concept of universal spontaneous
collapse of the wave function was conceived it was an attempt
to alter standard non-relativistic quantum physics. As such, it was
largely ignored by relativistic field theory and quantum gravity
communities. A central motivation of spontaneous collapse community
has been to replace the standard collapse-by-measurement that annoyed
many. With few exceptions, it did not annoy the field theory and quantum gravity
communities. Concept of certain general-relativity-related universal 
irreversibility in quantum field theory had been initiated very long ago 
by Wheeler, Hawking and a few others independently from the concept
of spontaneous collapse. Lately the two concepts started to converge and
support each other.
\end{abstract}

\section{Introduction}
Allow the author to introduce the topics in a self-quotation from 2015 \cite{Erice2015}.
`The inception of a universal gravity-related irreversibility took place originally in quantum cosmology but it turned out soon that a universal non-unitary dynamics is problematic itself. Independent investigations of the quantum measurement postulate clarified that a non-unitary dynamics is of interest already in the non-relativistic context. An intricate relationship between Newton gravity and quantized bulk matter might result in universal non-relativistic violation of unitarity - also called spontaneous decoherence. The corresponding gravity-related spontaneous decoherence model is now on the verge of detectability in optomechanical experiments. It is also a toy-model of cosmic quantum gravitational non-unitarity, illuminating that the bottle-neck of quantum gravity is the quantum measurement postulate instead of quantum cosmology.'

We want to build our message around the appearance of the concept of 
fundamental irreversibility in two separate fields of foundational theoretical 
physics. One is quantum cosmology. It has no consistent theory valid
down towards the Planck scale, despite various efforts using various advanced
mathematical models. At the same time, semiclassical and heuristic
approaches imply that fundamental irreversibilities may be part of quantum
cosmology. But quantum irreversibility itself, especially when universal,
happens to be non-trivial both technically and conceptually. And here
another research field gains importance. Its tasks seem opaque compared
to quantum cosmology's.  They include, e.g.,  the measurement problem,
the dynamics of wave function collapse, the quantum-classical transition 
and hybrid dynamics, but the relevant task is the quantum theory of macroscopia, 
coined as the \Schr~cat problem. You should discuss the options of
macroobjects' quantum behavior, and you can do it non-relativistically,
before you go for quantum cosmology! The hypothesis of spontaneous
universal collapse in massive degrees of freedom, assuming
a fundamental gravity-related irreverisibility, may have perspectives for
quantum cosmology as well, especially if fundamental irreversibility
remains part of it.

\section{Fundamental irreversibility? --- Two Communities}
The two fundamental issues in question are the following: unified theory
of space-time with quantized matter (cf. quantum gravity, quantum cosmology)
and the physics of quantum measurement (cf. wave function collapse, 
quantum-classical transition). These problems were considered unrelated for long time, 
studied by two separate research communities. 
We refer to the first one as quantum cosmologists and to other as 
measurement problem solvers or, with a spectacular touch, as \Schr~cat killers. 
Quantum cosmology has always been part of main stream physics, 
using heavy artillery of mathematics, while \Schr~cat killers used light weapons 
and sometimes whimsical identification of their problems. 
\begin{table}
\begin{center}
\caption{\label{T1}
Comparison of the two research fields and the  status of fundamental irreversibility.}
\begin{tabular}{|c|c|}
\hline
QUANTUM COSMOLOGISTS  &SCHR\"ODINGER CAT KILLERS$^*$\\
\hline
field-string-membrane theorists,&quantum foundation experts,\\
for full relativistic&for non-relativistic\\ 
quantum gravity&spontaneous wave function collapse\\
of the Universe&of macroscopic bodies\\
within standard quantum theory&with modified quantum theory\\
unitary (reversible)&non-unitary (irreversible)\\
\hline
Fundamental Irreversibility?&Fundamental Irreversibility?\\
Possible: foamy vacuum, black holes&Mandatory: wave function collapse\\
\hline
\end{tabular}
\end{center}
\small{$^*\!\!$ Measurement Problem Solvers}
\end{table}

\subsection{Irreversible Quantum Gravity/Cosmology at the Planck Scale}
Emergence of irreversibility \emph{within standard physics}  
was not derived by exact derivations but by \emph{heuristic arguments}.
A selection of earliest theoretical signatures could be the  following. 
In 1936, Bronstein \cite{Bro36} was the first to point out that space-time
metric must have an ambiguity $\delta g_{ab}$ of different nature from
the dictum of Heisenberg uncertainty relationships. 
Much later in 1962, Wheeler \cite{Whe62} proposed that space-time has a certain
foamy structure at the Planckian scale. After another decade, 
Bekenstein \cite{Bek73} proved that black-holes behave 
thermodynamically, they have entropy:
\begin{equation}
S_{BH}=\frac{A}{4},
\end{equation}
where $A$ is the surface aera of the black-hole in Planck units 
(Boltzmann's constant $k_B=1$). 
Then Hawking \cite{Haw75} showed that, yes, in accordance with the above entropy, 
black-holes emit thermal radiation of temperature
\begin{equation}
T_{BH}=\frac{1}{8\pi M},
\end{equation}
where $M$ is the black-hole mass. Later, Hawking \cite{Haw83}
went deeper and conjectured that unitarity of standard quantum theory is lost due to  
the instanton mechanism. The scattering processes must be non-unitary,
the non-unitary super-scattering operator $\$$ acts on the density operator $\ro$:
\begin{equation}
\ro\rightarrow\$ \ro
\end{equation}
where the r.h.s. does not factorize into ${\widehat S}\ro {\widehat S}^\dagger$
with any unitary scattering operator $\widehat S$.
Instead of Hawking's scattering process,
Banks, Susskind and Peskin \cite{BPS84} considered a  detailed dynamical equation
\begin{equation}
\frac{d\ro}{dt}=-i[\Ho,\ro]-\int\!\!\int [\Qo(x),[\Qo(y),\ro~]]h(x-y)d^3xd^3y,
\end{equation}
where $\Ho$ is the Hamiltonian,
$\Qo$ is a relativistic quantum field and $h$ is a positive kernel.
The authors show that the irreversible term on the r.h.s. violates the
energy-momentum conservation which cannot be restored in local
field theories.

\subsection{Irreversible Quantum Mechanics for Massive Objects}
Among diverse incentives, including a metaphysical discontent about von Neumann's quantum measurement theory,  
the only relevant motivation is the problematic extendibility of 
quantum theory for massive degrees of freedom. 
This led to \emph{heuristic modifications of standard
quantum physics}. According to that, the unitarity is violated in massive 
degrees of freedom so that macroscopically different superpositions,
aka \Schr~cat states like
\begin{equation}
|\Psi\rangle=\frac{|f_1\rangle+|f_2\rangle}{\sqrt{2}},
\end{equation}
collapse spontaneously into one of the components. 
Milestones of the consept could be the following.

In 1966, K\'arolyh\'azy \cite{Kar66}  proposed that, due to a conjectured spectrum of
space-time metric fluctuations $\delta g_{ab}$ at the Planck scale,
\Schr~cat states collapse before they get too large. He outlined a naive 
qualitative model of when and how the collapse happens.  
Independently and much later in 1986,
Ghirardi, Rimini and Weber \cite{GRW86} constructed a simple  model 
of universal spontaneous collapse in exact mathematical terms. 
They postulated extreme rare and weak 
spontaneous collapses for the elementary constituents' wave functions
which in turn yield robust localization of the macrosopic
center-of-mass, i.e., the collapse of the \Schr~cat state. The GRW model has
no reference to gravity. 
However, the present author  \cite{Dio87,Dio89} proposed  gravity-related universal 
spontaneous collapse of massive superpositions, like \Schr~cats, 
based on a master equation:
\begin{equation}
\frac{d\ro}{dt}=-\frac{i}{\hbar}[\Ho,\ro]-\frac{G}{2\hbar}\!\!\int\!\!\!\int [\fo(x),[\fo(y),\ro~]]\frac{1}{\vert x-y\vert}d^3xd^3y,
\end{equation}
where $\fo$ is the non-relativistic quantized mass density field and $G$ is
Newton's constant. The master equation corresponds to a certain ambiguity $\delta\Phi=\tfrac12 c^2 \delta g_{00}$ of the Newton potential. 

%Penrose (1996)  introduced a decay rate
%of \Schr~Cat 
%\begin{equation}
%\frac{G}{\hbar}\!\!\int\!\!\!\int [f_1(x)-f_2(x)][f_1(y)-f_2(y)]\frac{1}{\vert x-y\vert}d^3xd^3y
%\end{equation}
%where $f_1,f_2$ are the superposed mass densities of the Cat state.
%The master equation corresponds to a certain ambiguity $\delta\Phi$ of
%the Newton potential, Penrose's decay rate corresponds to the 
%related ambiguity of time-flow. 

\subsection{Fundamental irreversibility? --- Parallel pursuits}
Thus, the desired quantum gravity and the unwanted \Schr~cats
led both the heavy-armed relativistic and light-horse non-relativistic
studies to the same structure of heuristic master equations.
And they suffer of the same problem of spontaneous creation of 
energy and momentum. This coincidence is a spectacular instance
to illustrate that the two, apparently distant theoretical tasks, share
similar proposals and problems. Another instance has been the surprising 
coincidence between
the \Schr~cat's collapse rate predicted by the non-relativistic master 
equation \cite{Dio87} 
and by the general relativistic arguments of Penrose \cite{Pen96}.
For possible reasons of coincidence, see ref. \cite{Dio22}.
   
Mention should be made of Gell-Mann and Hartle \cite{GelHar90} 
who were perhaps the only ones at the time to recognize that quantum
cosmology needed a more general measurement theory than von Neumann's.
For the relationship of their theory of decoherent histories to spontaneous
collapse theories, see \cite{Dioetal95}. 
Most recently, hybrid quantum-classical master equations have been
tested in semiclassical cosmology by Oppenheim et al.  \cite{Opp22},
with numerous references to spontaneous collapse models which are, in fact, the
alternative formalism, see \cite{Dio23}.     

\begin{table}[h]
\begin{center}
\caption{\label{T2} 
A selection of parallel concepts related to fundamental irreversibility.}
\begin{tabular}{|l|l|l|}
\hline
           &QUANTUM COSMOLOGISTS  &SCHR\"ODINGER CAT KILLERS\\
\hline
1936&Bronstein \cite{Bro36}:  ambiguity $\delta g_{ab}$
          &\\
1962&Wheeler \cite{Whe62}: space-time foam
          &\\
1966&                                                         
         &K\'arolyh. \cite{Kar66}:  ambiguity $\delta g_{ab}$ collapses $\Psi$\\
1973&Bekenstein \cite{Bek73}: black hole entropy
         &\\ 
1975&Hawking \cite{Haw75}: black holes radiate
         &\\
1976&                                                         
          &Pearle \cite{Pea76}: searching collapse dynamics\\                        
1983&Hawking \cite{Haw83}: instantons break unitarity%$\rho_f=\$\rho_i\neq S\rho S^\dagger$
          &\\
1984&BPS \cite{BPS84}: energy non-conservation
          &Gisin \cite{Gis84}: prototype collapse dynamics\\ 
1986&                                                                                                               
          &GRW \cite{GRW86}: universal collapse\\
1987&
          &D. \cite{Dio87}: G-related decoherence\\            
1989&Ellis \& al. \cite{EllMohNan89}: wormholes collapse $\Psi$
          &D. \cite{Dio89}: G-related  collapse\\
1990&G-M\&H \cite{GelHar90}: decoherent histories
          &GRP \cite{GRP90}: G-unrelated collapse\\
1996&Penrose \cite{Pen96}: ambiguity $\delta g_{ab}$ collapses $\Psi$&\\
2014&
          &Bedingham \& al. \cite{Bedetal14}: relativisation\\
2016&Kwon\&Hogan \cite{KwoHog16}: holographic noise
          &Tilloy\&D \cite{TilDio16}: semiclassical G \& collapse\\
2017&
         &Bassi \& al. \cite{Basetal17}: review of G-collapse\\ 
2021&Sudarsky \cite{Sud21}: cosmology \& collapse
          &\\
2022&Anastopoulos\&Hu \cite{AnaHu22}: G-decoherence
          &\\
          &Oppenheim \& al. \cite{Opp22}: hybrid dynamics&\\
\hline
\end{tabular}
\end{center}
\end{table}

\section{Summary}
The failures of quantum gravity models may not be due to the inadequacy
of the quantization methods tried so far, but the inadequacy of standard
quantum theory for macroobjects, e.g., the spontaneous decay of 
massive superpositions, aka \Schr-cats. We recalled that the idea of
fundamental irreversibility has also been raised in quantum cosmology. 
The roots of irreversibility may be common in
quantum cosmology and in speculative models of spontaneous
collapse of the wave function. This may be so despite the fact
that one is predicted at the Planck scale while the other comes from
non-relativistic discussion of massive superpositions \cite{Dio18}.      
Considerations of  quantum cosmologists and advocates of 
spontaneous collapse (referred here extravagantly as \Schr~cat killers) 
will have to come closer to each other. To support this suggestion,
here are two quotes. The author (cat killer) claimed 
`the measurement problem culminates in quantum cosmology" \cite{Dio92}.
The cosmologist claims  `the principles of general relativity
must influence, and actually change, the very formalism of quantum mechanics'
\cite{Pen14}.

\ack
This research was funded by the Foundational Questions Institute and Fetzer Franklin
Fund, a donor-advised fund of the Silicon Valley Community Foundation (Grant No's. FQXi-RFPCPW-2008,  FQXi-MGA-2103), the National Research, Development and Innovation Office (Hungary)
``Frontline'' Research Excellence Program (Grant No. KKP133827),
and the John Templeton Foundation (Grant 62099).

\end{document}